\documentclass[twocolumn,aps,prl,showpacs,raggedbottom,nobalancelastpage,amssymb,superscriptaddress]{revtex4-1}

\usepackage{graphicx}
\usepackage{grffile}
\usepackage{amsmath}
\usepackage{amssymb}
\usepackage{bm}
\usepackage[usenames]{color}
\usepackage{amsfonts}
\usepackage{appendix}
\usepackage{dsfont}

\def \dH{d^{\text{H}}_{n}}

\def \dF{d^{\text{F}}_{n}}

\begin{document}
\title{Observation of Topological Phase Transitions in Photonic Quasicrystals}
\author{Mor Verbin}
\affiliation{Department of Physics of Complex Systems, Weizmann Institute of Science, Rehovot 76100, Israel.}
\author{Oded Zilberberg}
\affiliation{Department of Condensed Matter Physics, Weizmann Institute of Science, Rehovot 76100, Israel.}
\author{Yaacov E.~Kraus}
\affiliation{Department of Condensed Matter Physics, Weizmann Institute of Science, Rehovot 76100, Israel.}
\author{Yoav Lahini}
\affiliation{Department of Physics of Complex Systems, Weizmann Institute of Science, Rehovot 76100, Israel.}
\affiliation{Department of Physics, Massachusetts Institute of Technology, Cambridge MA, USA.}
\author{Yaron Silberberg}
\affiliation{Department of Physics of Complex Systems, Weizmann Institute of Science, Rehovot 76100, Israel.}

\begin{abstract}
Topological insulators and topological superconductors are distinguished by their bulk phase transitions
and gapless states at a sharp boundary with the vacuum. Quasicrystals have recently been found to be topologically nontrivial. In quasicrystals, the bulk phase transitions occur in the same manner as standard
topological materials, but their boundary phenomena are more subtle. In this Letter we directly observe
bulk phase transitions, using photonic quasicrystals, by constructing a smooth boundary between
topologically distinct one-dimensional quasicrystals. Moreover, we use the same method to experimentally
confirm the topological equivalence between the Harper and Fibonacci quasicrystals.\end{abstract}


\maketitle

The classification of gapped systems, such as band
insulators and superconductors, by topological indices
is a rapidly developing paradigm in condensed matter
physics \cite{TI_RMP1}. This novel approach provides insights into
the characterization of states of matter, as well as predicts
exotic phenomena.

The topological classification of a system assigns an
integer index to its energy gap. This index encodes properties
that are robust to distortions and deformation of
the system. Hence, when a system with a given index is
continuously deformed into a system whose index has a
different value, the bulk energy gap must close, namely a
quantum phase transition occurs. Accordingly, if a system
can be continuously deformed into another system while
keeping the bulk gap open, then their topological indices
must be the same, defining them as topologically equivalent.
Usually, such a phase transition manifests by the
appearance of gap-traversing states at a sharp boundary
between a topologically nontrivial material and the topologically
trivial vacuum. Examples for such boundary
states are the chiral modes of the integer quantum Hall
effect (IQHE), the Dirac cone of the three-dimensional
topological insulator, and the Majorana fermions of one-dimensional
(1D) topological superconductors \cite{TI_RMP1}.

A new type of topological phenomena has been recently
studied in quasiperiodic systems \cite{Original_Recipe,Smoothening}. Such systems,
which are ordered but not periodic, were shown to be
characterized by topological indices that are usually attributed
to systems of a dimension higher than their physical
dimension. In particular, it was shown that the canonical
1D quasiperiodic systems, i.e., the Harper (or Aubry-Andr\'{e}) model \cite{Harper,AA}, the diagonal Fibonacci model \cite{KKT}, and their quasicrystalline off-diagonal variants \cite{Han,oDiagFib}, can
be assigned Chern numbers. These numbers are topological
indices that characterize generic two-dimensional (2D)
systems. A continuous deformation between two quasicrystals
(QCs) with different Chern numbers will therefore
result in a bulk phase transition. Correspondingly, at a
sharp boundary between such a QC and the vacuum,
localized subgap boundary states may appear. These
states were experimentally observed in photonic Harper
QCs \cite{Original_Recipe}. Further analyses of 1D topological QCs with sharp boundaries have been conducted in cold atoms
\cite{citing_us2} and superconducting wires \cite{Tezuka:2012}.

In systems where the topology is generated by a
symmetry which is broken at the boundary, the boundary
phenomenon is not robust \cite{WTI,CTI}. Similarly, the topological
characterization of a QC is based on its long-range
order, which is broken at the sharp boundary. Therefore,
its boundary states do not always appear. However, if the
boundary between a topologically nontrivial QC and a
topologically trivial system is adiabatically smooth, subgap
states will always appear at the boundary, revealing the bulk
gap closure. Accordingly, a smooth boundary can be used to
prove equivalence between different quasiperiodic systems,
if the energy gap remains open throughout the deformation.

In this Letter, we study the bulk gap closure that occurs
when smoothly deforming between topologically inequivalent
quasiperiodic systems, and its absence when the
systems are topologically equivalent. To this end, we create
two inequivalent Harper QCs, and spatially deform
between them. We then observe the closure of bulk energy
gaps through the emergence of subgap states localized
within the deformation region. In contrast, using the
same interpolation process between seemingly different
but topologically equivalent systems, the Harper and the
Fibonacci QCs, no such phase transition is observed,
thereby confirming that these two models are indeed topologically
equivalent. These phenomena are experimentally
tested in quasiperiodic photonic lattices, where the phase
transition, or its absence, is directly observed using the
propagation of light in waveguide arrays.

Photonic lattices are widely used for realizations of
different models originating from solid state physics, due to the high level of control over their parameters and
behavior \cite{Waveguides_Review2}.
Our photonic QCs are composed of an
array of coupled single-mode waveguides, fabricated in
bulk glass using femtosecond laser microfabrication technology \cite{Waveguide_Fabrication2}. The overlap between the evanescent modes
of the waveguides allows the propagating light to tunnel
from each waveguide to its neighboring waveguides.
Hence, the hopping amplitude between adjacent waveguides
can be controlled by modulating the spacing
between them.

The dynamics of light propagating in these coupled
waveguide arrays is described by the tight-binding model,
with the propagation axis $z$ taking over the role of time, $i\partial_{z}\psi_{n}=H\psi_{n}$, where $\psi_{n}$ is the wavefunction at waveguide number $n$. Taking the hopping amplitude to be real,
we obtain the general Hamiltonian
\begin{align}
H\psi_{n} & =t_{n}\psi_{n-1}+t_{n+1}\psi_{n+1} \, ,\label{Eq:H_TBM}
\end{align}
where $t_{n}$ is the hopping amplitude from site $n$ to site $n-1$.

\begin{figure}
\includegraphics[width=3.38in,clip]{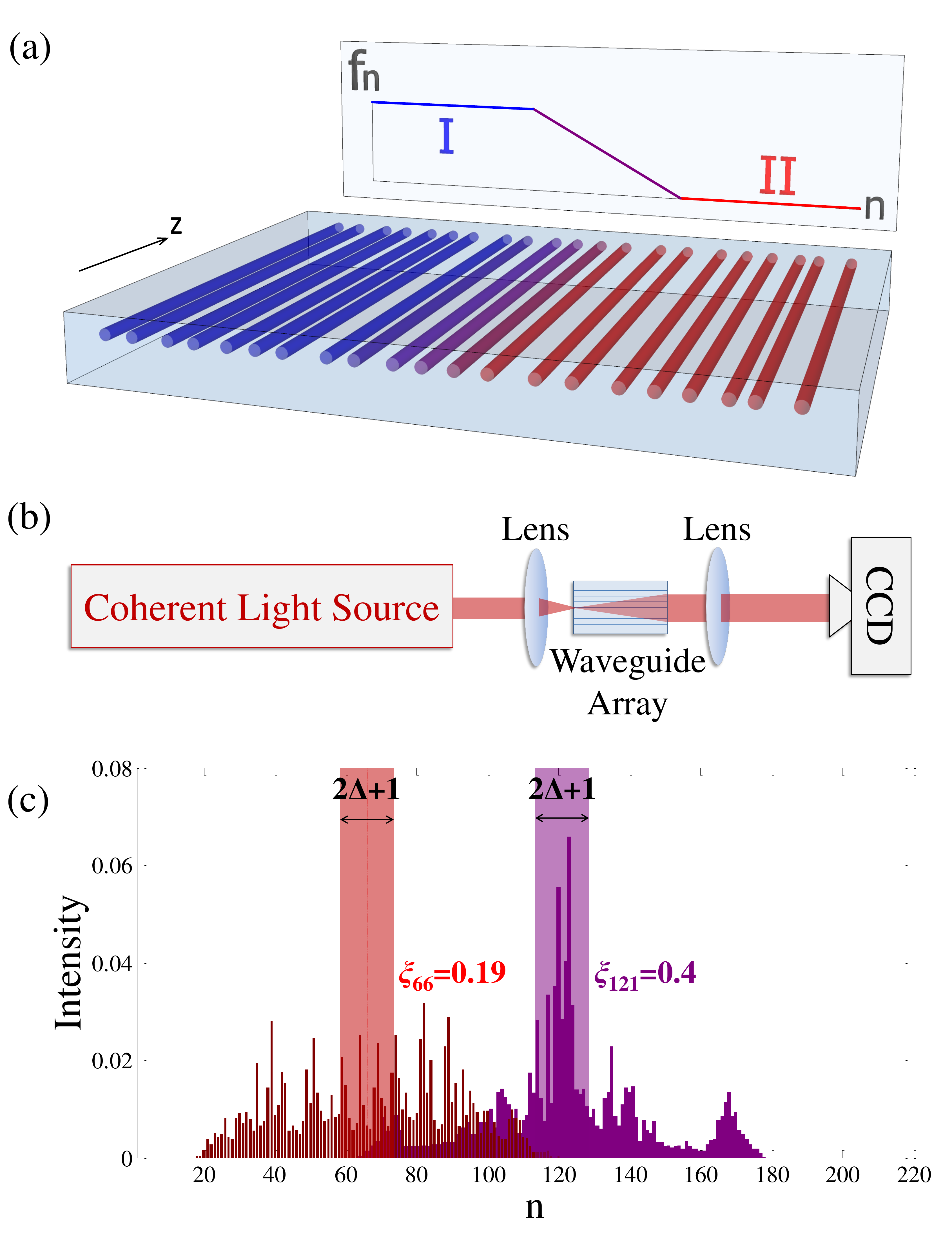}
\caption{\label{Experimental Methods} (Color online) Experimental methods. (a) Illustration of a photonic waveguide array implementing a deformation
between two QCs [cf.~Eq.~\eqref{Eq:Deformation}]. (b) A schematics of the experimental setup. We focus a coherent light beam into a waveguide
in the array, allow it to propagate along the structure, and image
the output intensity using a CCD camera. (c) Illustration of the
relation between the localization parameter $\xi_{n}$ and the measured intensity distribution $|\psi_n|^2$ for two injection sites, $n=66$ and $n=121$. The shaded intervals denote the $\Delta$-distance neighborhood around the insertion point. When more light remains within this neighborhood, the value of $\xi_n$ increases.}
\end{figure}

Our intention is to study the transition that occurs when
some system $I$ is deformed into another system $II$, where
each system has its own set of quasiperiodic hopping
amplitudes $t_{n}^{I}$ and $t_{n}^{II}$, respectively.
To this end, we fabricate
a waveguide array with a deformed hopping profile $t_{n}=f_{n}t_{n}^{I}+\left(1-f_{n}\right)t_{n}^{II}$, where
\begin{align} \label{Eq:Deformation}
\resizebox{.9\hsize}{!}{$\displaystyle f_{n} = \begin{cases}
\begin{array}{c}
1\\
1-\frac{n-L_{I}}{L_{D}}\\
0
\end{array} &
\begin{array}{c}
1\leq n\leq L_{I}\\
L_{I}< n<L_{I}+L_{D}\\
L_{I}+L_{D}\leq n\leq L_{I}+L_{D}+L_{II}
\end{array},
\end{cases}$}
\end{align}
as depicted in Fig.~\ref{Experimental Methods}(a). This procedure produces an array of length $L_{I}$ of system $I$ on one
side of the structure, an array of length $L_{II}$ of system $II$ on the other side, and an $L_{D}$-long deformation region,
which continuously transforms between the two. This structure enables the study
of the eigenstates of both systems as well as the transition
between them, on a single waveguide array.

The properties of the Hamiltonian fabricated within the
photonic crystal are studied by injecting light into one of
the waveguides in the array and measuring the outgoing
intensity at the output facet using a CCD camera, as
illustrated in Fig.~\ref{Experimental Methods}(b).
The injected beam excites a
wave packet of all the modes that have a nonvanishing
amplitude at the injection site, and the light propagates in
the lattice according to this superposition of eigenstates.
The width of the outgoing intensity distribution can therefore
reveal the existence of localized eigenstates: If there
is no localized state near the injection site, the light
spreads freely throughout the array, propagating according
to the bulk properties of the system. However, when
light is injected in the vicinity of a localized state, its
expansion is dominated by the width of the state. To
quantify the localization of the outgoing light, we measure
the amount of light that remains within a small
distance $\Delta$ from the injection site $n$, by measuring the generalized return probability \cite{xi},
\begin{equation} \label{Eq:xi_n}
 \xi_{n}=\displaystyle \Big( \sum\limits_{m={n-\Delta}}^{n+\Delta} |\psi_m|^2 \Big) / \Big( \sum\limits_{m=1}^{L_{I}+L_{D}+L_{II}} |\psi_m|^2 \Big) \, .
\end{equation}
The relation between $\xi_n$ and the intensity distribution $|\psi_n|^2$ is illustrated in Fig.~\ref{Experimental Methods}(c).
Since $\xi_n$ is meant to reveal the existence of localized states, we will choose $\Delta$ to be of the order of the width of a localized subgap state.

Let us now introduce the specific quasiperiodic tight-binding
models under study. In these models the hopping
amplitude is modulated according to
\begin{align} \label{Eq:Hopping}
t_n & =t_{0}\left[1+\lambda d_n\right] \, ,
\end{align}
where $t_{0}$ is the characteristic hopping amplitude of the system, $\lambda \in [0,1)$ is the modulation strength, and $d_n \in [-1,1]$ is some quasiperiodic modulation function.

Here we consider two such modulations: the Harper modulation
\begin{align} \label{Eq:Hopping_Harper}
\dH=\cos(2\pi bn+\phi) \, ,
\end{align}
and the Fibonacci-like modulation
\begin{equation} \label{Eq:d_Fibonacci}
\resizebox{.85\hsize}{!}{$\displaystyle \dF=2\left(\left\lfloor \frac{\tau}{\tau+1}\left(n+2\right)\right\rfloor -\left\lfloor
\frac{\tau}{\tau+1}\left(n+1\right)\right\rfloor \right)-1=\pm1 \, .$}
\end{equation}
The long-range order of the Harper QC \cite{Han} is controlled by the modulation frequency $b$. Whenever $b$ is irrational, the
hopping modulation is incommensurate with the underlying lattice, resulting in a quasiperiodic pattern. Accordingly, the
parameter $\phi$ shifts the origin of the modulation. Comparably, the Fibonacci-like QC is constructed
from a sequence of two values that are ordered in a quasiperiodic
manner. This sequence is obtained by applying the ``cut-and-project'' procedure on a square lattice onto the line $y=x/\tau$ \cite{QC_Senechal}. Whenever the slope of the line, $\tau$, is irrational, the sequence
is quasiperiodic. For example, the case of $\tau = (1+\sqrt{5})/2$ is the well-known Fibonacci QC.

The energy spectrum of the Harper QC is composed of a
fractal set of bands and gaps, in a way that depends on the
modulation frequency $b$ \cite{Hofstadter,Han}. These gaps are associated
with Chern numbers, which are also uniquely determined
by $b$ \cite{Smoothening}. For any rational approximant $b=p/q$, the Chern number $\nu_r$ that is associated with a gap number
$r=1,\ldots,(q-1)$ abides the Diophantine equation $r = \nu_r q + t_r p$, where $\nu_r$ and $t_r$ are integers, and $0 < |\nu_r|
< q/2$ \cite{TKNN}. The distribution of Chern numbers for an irrational $b$ is given by taking the appropriate limit of $p,q
\rightarrow \infty$. Hence, the gaps of two Harper QCs with $b_{I}\neq b_{II}$ are associated with different distributions of
Chern numbers. Thus, when deforming between two such models the
Chern number distribution rearranges by level crossings.
This makes these models topologically inequivalent.

The properties of the Fibonacci-like QC differ in many
ways from those of the Harper QC, e.g., the localization of
the bulk wave functions \cite{KKT,oDiagFib,Han,Svetlana}. Nevertheless, it was
recently shown that they are topologically equivalent
whenever the frequency of the Harper modulation satisfies $b=(\tau+1)/\tau$ \cite{Smoothening}. In such a case, the gaps of the
Fibonacci-like QC are associated with the same Chern
numbers as those of the Harper QC. Hence, for a given
modulation frequency b, the Harper QC can be continuously
deformed into the Fibonacci-like QC without the
appearance of a phase transition.

Note that the deformation in Eq.~\eqref{Eq:Hopping_Harper} contains an addition degree of freedom in the form of the parameter $\phi$. This
parameter has a crucial role in the observation of the
topological boundary states of quasiperiodic systems \cite{Original_Recipe}.
While the spectrum of our model is gapped in the bulk,
localized boundary states appear, which traverse the energy
gaps as a function of $\phi$. Nevertheless, in this Letter we
focus on bulk properties, which are $\phi$-independent \cite{SUPMAT}.

\begin{figure*}[htb]
\includegraphics[width=\textwidth]{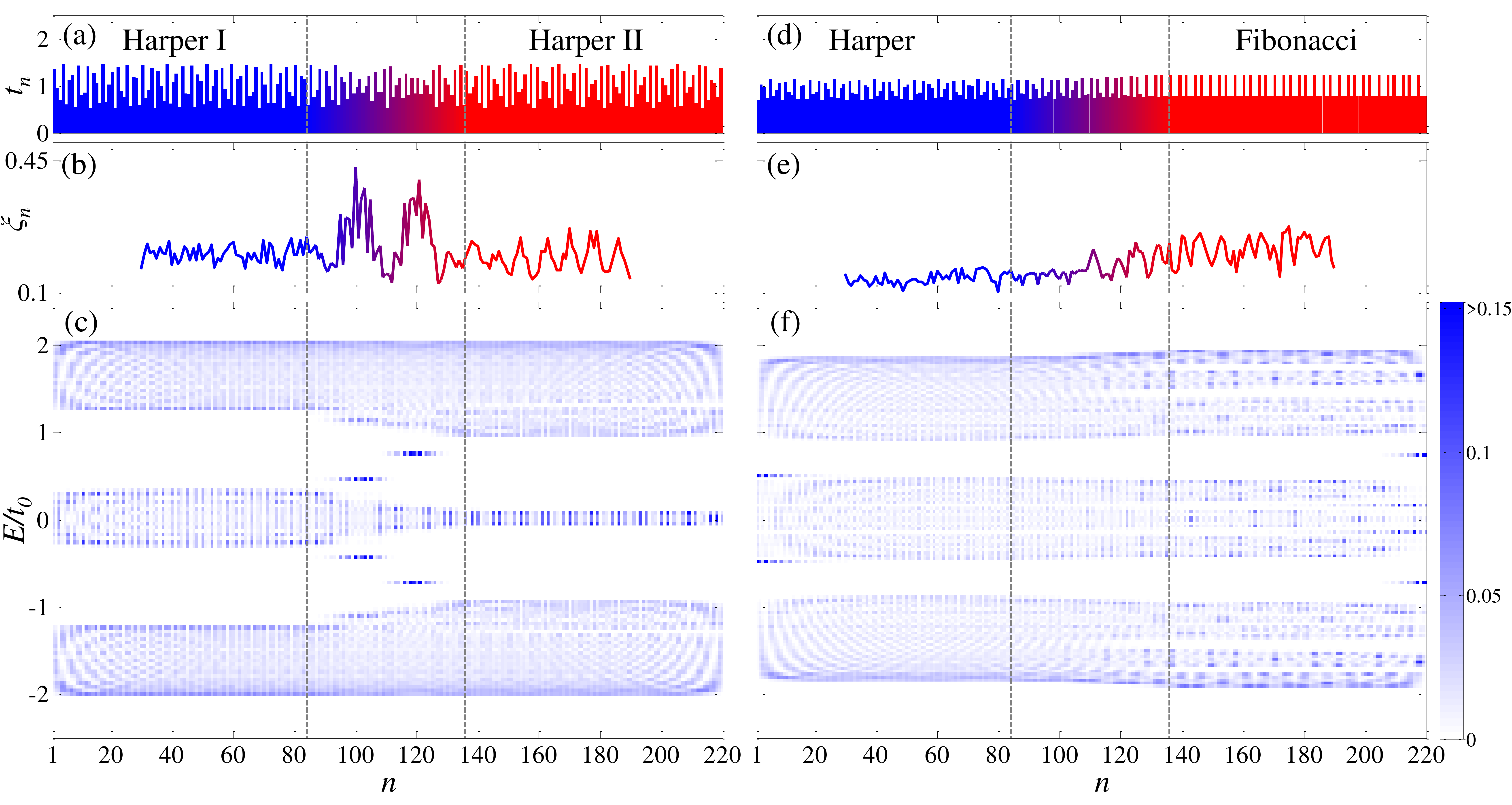}
\caption {\label{Phase and No Phase Transition} (Color online) Summary of results for a smooth deformation between (a)–(c) two topologically inequivalent Harper QCs, and
(d)–(f) topologically equivalent Harper and Fibonacci QCs. In both experiments (from left to right), $L_{I}=84$ (blue), $L_{D}=51$ (purple hues), and $L_{II}=84$ (red).
(a) The hopping amplitude $t_{n}$ as a function of the lattice site $n$, for modulation frequencies $b_{I}=2/(1+\sqrt{5})$ and $b_{II}=2/(1+\sqrt{6.5})$ of the Harper QCs.
(b) Experimentally-measured $\xi_{n}$ for $\Delta=7$, as a function of the injection site $n$. The two peaks within the deformation region imply the existence of localized states.
(c) Numerically obtained LDOS of the structure, $D_{n}(E)$. The energy bands are composed of extended states, while localized states (roughly 15 sites wide) traverse the gaps in the deformation
region. These states manifest the transition between the inequivalent QCs. (d)–(f) Same as (a)–(c), but with a Harper QC deformed into
a Fibonacci QC with $b_I=(\tau_{II}+1)/\tau_{II}=2/(1+\sqrt{5})$.
Here, $\xi_{n}$ shows no sign of localized states. Accordingly, though the distribution of the bands changes along the structure, the energy gaps appear to remain open. This confirms the equivalence between the two QCs.}
\end{figure*}

We now turn to our experimental results. We construct a
deformation between topologically inequivalent Harper
QCs with modulation frequencies $b_{I} \neq b_{II}$. Figure \ref{Phase and No Phase Transition}(a) depicts the hopping amplitudes of a deformation between
a Harper QC with $b_{I}=2/(1+\sqrt{5})$ and a Harper QC with $b_{II}=2/(1+\sqrt{6.5})$, where $t_{0}=28/75$ mm$^{-1}$, $\lambda=0.475$ and
$\phi_I=\phi_{II}=\pi$.
For this set of parameters, the bulk
wave functions of the Harper QCs are extended \cite{Han}.
We fabricated this structure in a $75$mm-long photonic waveguide array. This results in an effective propagation of $14$ tunneling lengths, where the tunneling length is the characteristic length for hopping, namely $2/t_{0}$. With such a propagation length, light injected in the bulk of the structure
will sufficiently expand in comparison to the width of
localized subgap states.

To experimentally observe the phase transition between the two QCs, a 808nm continuous-wave diode laser beam was injected into each waveguide with a $\times 40$ microscope objective. The light at the output facet was imaged onto a CCD camera using a $\times 5$ microscope objective.
Using the measured light distribution, we obtained $\xi_n$ as a function of the injection site $n$.
The results are presented in Fig.~\ref{Phase and No Phase Transition}(b).
Two clear peaks in the deformation region can be seen over a relatively flat $\xi_n$ outside the region.
This is a clear indication of the existence of localized states within the deformation region.
Note that measurements of $\xi_n$ for $n<30$ and $n>190$ are omitted from this plot---for these injection sites the expanding light hits the edges of the structure, causing $\xi_n$ to be skewed by boundary effects.

To reveal the source of the peaks observed in $\xi_n$, we numerically obtain the local density of states (LDOS), which is
presented in Fig.~\ref{Phase and No Phase Transition}(c). The LDOS is defined by
$D_{n}\left(E\right)=\underset{m}{\sum}\delta\left(E-E_{m}\right)\left|\varphi_{n}^{(m)}\right|^{2}$, where $E_{m}$ is the energy
of the $m^{\rm th}$ eigenstates, and $\varphi_{n}^{(m)}$ is its wavefunction. $D_{n}(E)$ describes the spatial distribution of
the eigenstates of the structure as a function of energy. For $n\leq L_{I}$, we observe bands of extended states that correspond
to the eigenstates of system $I$. Similarly, for $n\geq L_{I}+L_{D}$ we recognize the band of extended states of system $II$.
However, along the deformation region, there are few spatially localized states
with energies that discretely traverse the gaps. These subgap
states are the origin of the measured peaks in $\xi_n$. Their
appearance is an explicit signature of the phase transition
between the inequivalent QCs since they traverse the gap
continuously when $L_D \rightarrow \infty$ \cite{SUPMAT}. We have therefore experimentally observed the bulk phase transition between
two topologically inequivalent Harper QCs.

We now turn to study the transition between topologically
equivalent QCs. We constructed a deformation
between a Harper QC and Fibonacci QC, with a matched
modulation frequency $b_I=(\tau_{II}+1)/\tau_{II}=2/(1+\sqrt{5})$ \cite{shift}. The hopping
amplitudes $t_n$ are depicted in Fig.~\ref{Phase and No Phase Transition}(d), for $t_{0}=28/75$ mm$^{-1}$, $\lambda=0.225$,
$\phi=\pi(1+3b)$. For this set of parameters, the bulk wave functions of the
Harper QC are extended, while those of the Fibonacci are critical \cite{oDiagFib}. Nevertheless, for the structure's effective
propagation length, light injected into the bulk of both QCs
will sufficiently expand in comparison to the width of
potential localized states. The measured $\xi_n$ of this system is depicted in Fig.~\ref{Phase and No Phase Transition}(e), showing no sign of
localized states within the deformation region. The numerically obtained LDOS is shown in Fig.~\ref{Phase and No Phase
Transition}(f). While the configuration of the
bands changes considerably between the two QCs, no gap
closure is observed along the deformation. Note, also,
that two subgap states appear at the sharp boundaries
with the vacuum \cite{SUPMAT}. The open gaps and the corresponding
absence of peaks in $\xi_n$ serve as experimental confirmation
of the equivalence between the Fibonacci and the
Harper QCs.

To conclude, in this Letter we have presented a novel
method to study topological phase transitions using a
continuous deformation between two systems, which acts
as a smooth boundary between them. When the boundary is
sufficiently smooth, observations of subgap states localized
within the deformation area serve as evidence of the
phase transition. Such subgap states do not appear when a
phase transition does not take place, namely, between
topologically equivalent systems. Our method extends the
prevailing approach which focuses on states that appear at
sharp boundaries between topologically nontrivial systems
and the vacuum. It has proven useful in studying the
topological characterization of 1D quasiperiodic systems,
since it circumvents the subtlety of their boundary phenomena
at sharp boundaries. Furthermore, this technique
may be useful to study other topological systems, such as
(i) the weak and the crystalline topological insulators,
where the surface breaks the underlying symmetry \cite{WTI,CTI}, (ii) varying dopant concentration in 3D topological
insulators \cite{Sato:2011}, and (iii) nanowires that may host
Majorana fermions at their boundaries \cite{Mourik:2012}.

We thank Y.~Avron, R.~Lifshitz, A.~Keselman and S.~Huber for fruitful discussions. We thank the U.S.~Israel Binational Science
Foundation, the Minerva Foundation of the DFG, Crown Photonics Center, ERC Advanced Grant Quami, and ISF Grant No. 700822030182 for financial support.

\newpage
\begin{center}
\textbf{\large SUPPLEMENTAL MATERIAL}
\end{center}
\vspace{5mm}

\section{I. Effect of finite-sized deformation on the phase transition}

\begin{figure}[b]
\includegraphics[width=8cm,height=9cm]{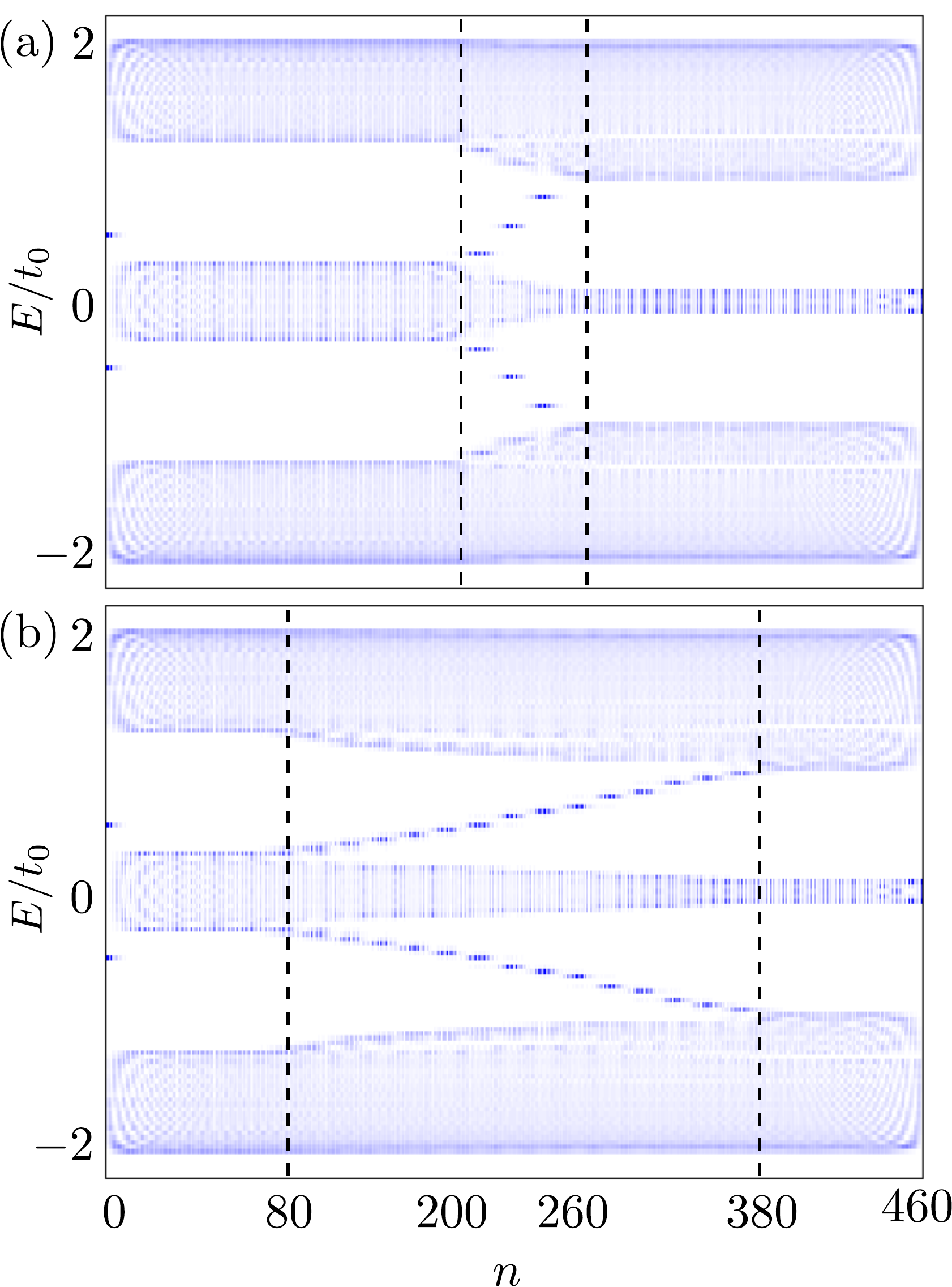}
\caption{\label{ldos} Numerically obtained LDOS for deformation between inequivalent Harper QCs, with $b_I=2/(1+\sqrt{5})$ and
$b_{II}=2/(1+\sqrt{6.5})$. (a) The deformation length is $L_D=60$. The gap is traversed by $3$ states in the deformation region.
(b) For $L_D = 300$, there are $15$ subgap states, and the level spacing between them decreases correspondingly. }
\end{figure}

In the main text, we study the phase transition that occurs when two topologically distinct quasicrystals (QCs) are deformed from
one into the other, and its absence when the two QCs are topologically equivalent. This transition can be obtained by
adiabatically deforming between the two QCs in real space, thereby creating a smooth boundary between them. As a result, when a
topological phase transition occurs, the energy gap is continuously traversed by states localized within the deformation region.
However, in physical systems, the deformation takes place over a finite length, making it not strictly adiabatic, and the gap
closure is disrupted. Nevertheless, a finite number of subgap states remain, and their energies discretely traverse the gap. Here
we show that these states serve as evidence of the topological phase transition, since their level spacing decreases with the
deformation length.

\begin{figure}[htb]
\includegraphics[width=7cm,height=4.5cm]{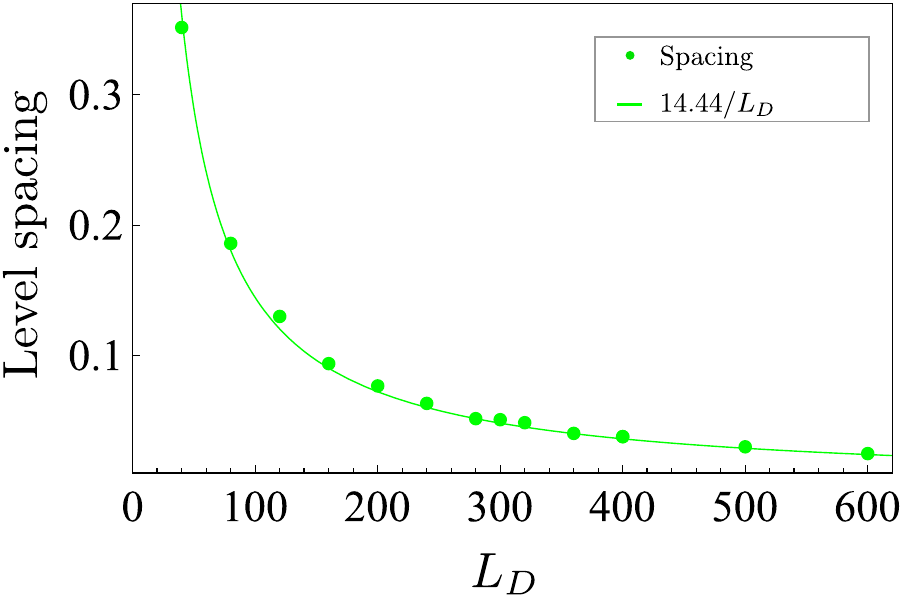}
\caption{\label{ld} Level spacing between two states in the middle of a large gap as a function of the deformation length $L_D$.
While the width of the states ($l_\text{sub}$) is fixed, the number of states scales as $L_D$, and their level spacing scales as
$l_\text{sub}/L_D$. The dots are numerically obtained data, and the fitted curve matches $14.44/L_D$.}
\end{figure}

Figure \ref{ldos} depicts the numerically obtained local density of states (LDOS) for structures deforming between two
topologically distinct Harper QCs. For this plot, the parameters are $b_{I}=2/(1+\sqrt{5})$, $b_{II}=2/(1+\sqrt{6.5})$,
$\lambda_I = \lambda_{II}=0.475$, $\phi_I = \phi_{II} = 0$, and $L_I+L_D+L_{II} = 460$. In Fig.~\ref{ldos}(a), the deformation
length is $L_D = 60$, and $3$ subgap states appear within each one of the large gaps. In Fig.~\ref{ldos}(b), $L_D = 300$, and
$15$ subgap states appear. As $L_D$ increases, the number of subgap states within a gap ($N_\text{sub}$) increases, while the
spacing between their energies (level spacing) decreases. Figure \ref{ld} plots the level spacing between two states residing in
the middle of a large energy gap as a function of $L_D$. We can see that the level spacing scales as $\sim 1/L_D$. Therefore, in
the limit of $L_D \rightarrow \infty$, we approach the continuous phase transition.

The level spacing can be approximated by $\Delta / N_\text{sub}$, where $\Delta$ is the size of the gap. The scaling of $1/L_D$
implies that there is an additional length scale, namely the width of the subgap states, $l_\text{sub}$. Given a fixed $l_\text{sub}$,
the number of subgap states within the deformation region is approximately $N_{\rm sub} \approx L_D/l_\text{sub}$, and their
level spacing scales like $l_\text{sub}/L_D$. The width $l_\text{sub}$ controls the smoothness of the boundary, and sets a lower
bound on the deformation length for which subgap states always appear, i.e.~$L_D > l_\text{sub}$. We can therefore conclude that
structures with a finite deformation length larger than $l_\text{sub}$ can indeed serve as a probe of the existence of a
topological phase transition.



\section{II. Boundary states at a sharp boundary and the role of $\phi$}

We have seen that subgap states appear within a finite-length boundary between topologically inequivalent QCs. These states
always appear for sufficiently long deformation regions ($L_D \gg l_{\rm sub}$), and their number increases with the deformation
length $L_D$.

\begin{figure}[b!]
\includegraphics[width=8.2cm]{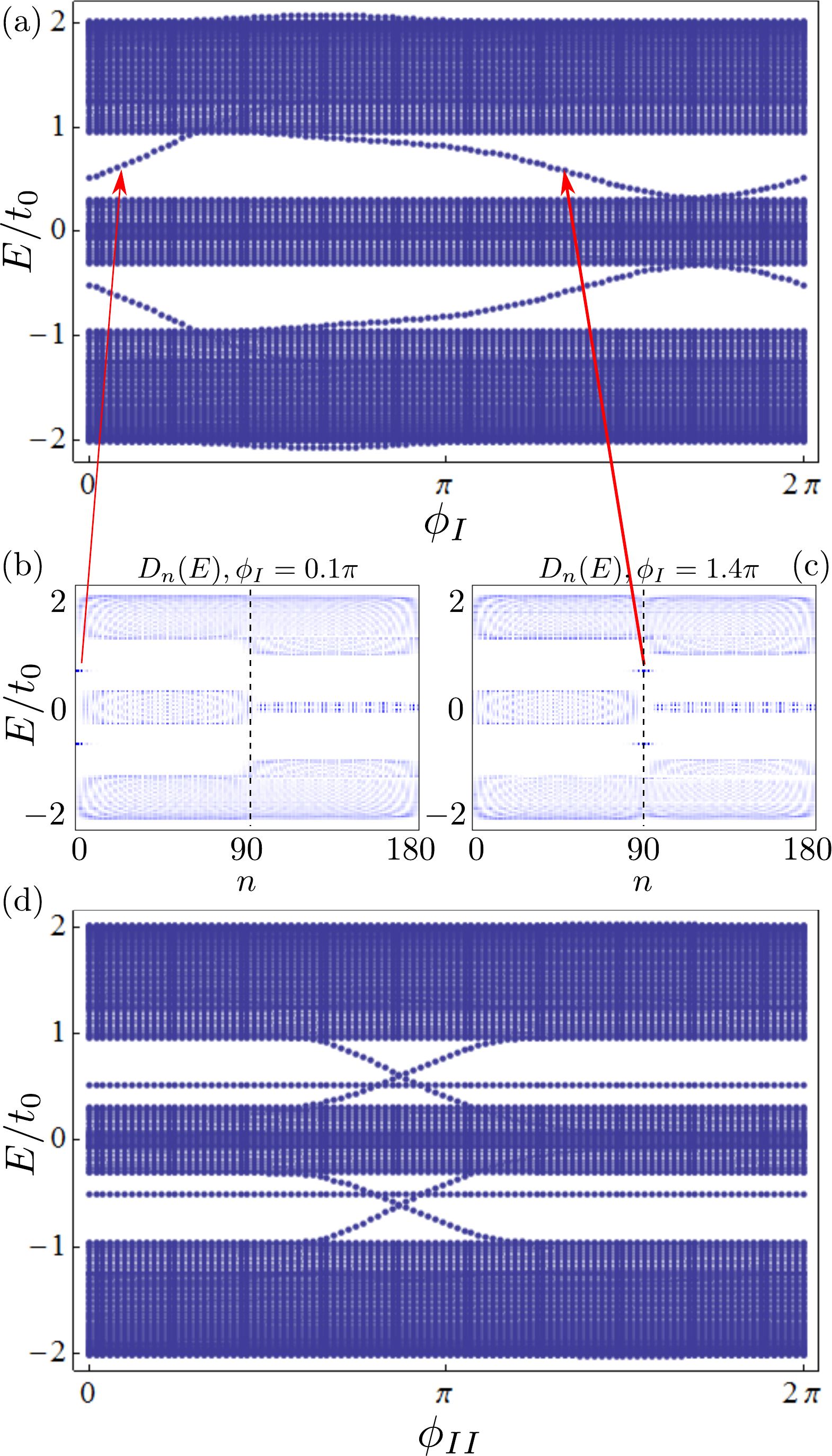}
\caption{\label{Fig3} Boundary states at a sharp boundary between two QCs. (a) Numerically obtained spectrum of a structure with
the same parameters as in Fig.~\ref{ldos} and $L_D = 0$ as a function of $\phi_I$. The large gaps are traversed by two states
when $\phi_I$ is scanned from $0$ to $2\pi$. (b) The corresponding LDOS for $\phi_I = 0.1\pi$. The gap-traversing state is
localized at the boundary of QC $I$ with the vacuum. (c) The corresponding LDOS for $\phi_I = 1.4\pi$. The subgap state is
localized at the sharp boundary between QC $I$ and QC $II$. (d) The spectrum of the structure as a function of $\phi_{II}$.
Again, two states traverse the gaps, where one of them is localized at the boundary of QC $II$. The horizontal line in the middle
of the gaps belongs the boundary state of QC $I$, which is unaffected by $\phi_{II}$. }
\end{figure}

In the opposite limit of $L_D\rightarrow 0$, the boundary becomes sharp. At a sharp boundary, subgap states may or may not
appear, depending on the exact pattern of the QCs~\cite{Original_Recipe}. Recall that the Harper modulation [cf.~Eq.~(5) in the
main text] includes the parameter $\phi$. This parameter shifts the origin of the quasiperiodic pattern, and is known as a
phasonic degree of freedom. The appearance of subgap states at a sharp boundary with the vacuum is controlled by $\phi$, since it
determines the exact pattern of the QC at the termination point. As $\phi$ varies from $0$ to $2\pi$, boundary states
continuously traverse the energy gaps. The number of gap-traversing states is given by the corresponding Chern number, which is
determined by the modulation frequency $b$.

At a sharp boundary between two QCs, subgap states traverse the gap as either $\phi_{I}$ or $\phi_{II}$ is scanned. This can be
seen in Figs.~\ref{Fig3}(a)-(d), which show the numerically obtained spectra of a structure with the parameters from the previous
section and $L_D = 0$, as a function of $\phi_I$ and $\phi_{II}$. As $\phi_I$ is scanned, two states traverse each one of the
large gaps, one localized at the sharp boundary of QC $I$ with the vacuum, and the other at the sharp boundary between QC $I$ and
QC $II$ [as seen from the corresponding LDOS in Figs.~\ref{Fig3}(b) and \ref{Fig3}(c)]. A similar behavior is observed when
$\phi_{II}$ is scanned, only now, one of two the gap-traversing states is localized at the sharp boundary of QC $II$ with the
vacuum.

\begin{figure}[ht!]
\includegraphics[width=8.1cm]{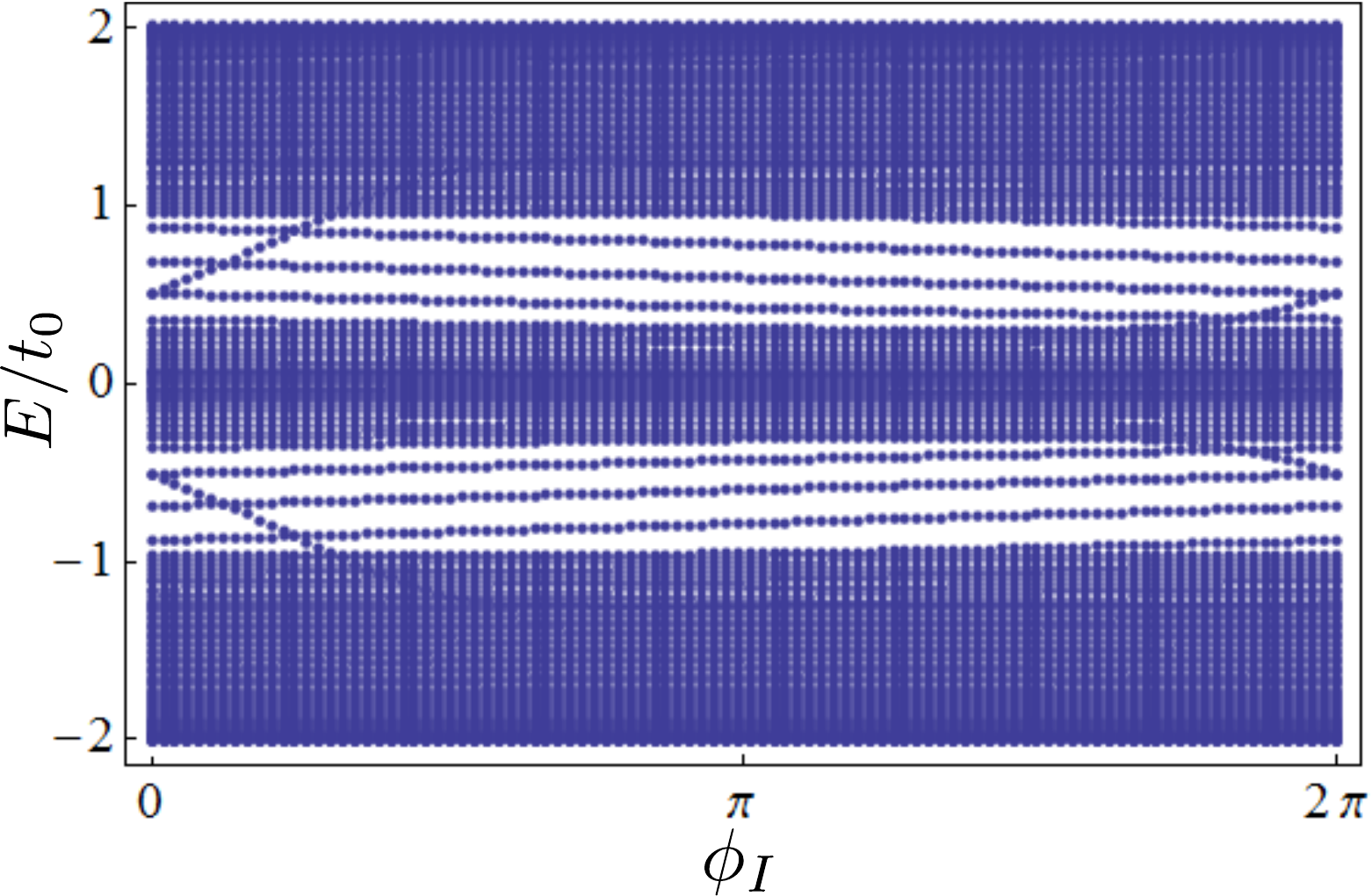}
\caption{\label{Fig4} Spectrum of a structure with the same parameters as in Fig.~\ref{ldos} and $L_D = 80$ as a function of
$\phi_I$. The energy of each subgap state in the deformation region approaches the energy of the next one when $\phi_I$ is
scanned. The state at the boundary of QC $I$ with the vacuum behaves like in Fig.~\ref{Fig3}(a). }
\end{figure}

In contrast to a sharp boundary, for infinitely long $L_D$, the gap traversing states are independent of $\phi$. For finite
$L_D$, where there is a finite number of subgap states, the gap completely closes when $\phi$ is scanned. This can be seen in
Fig.~\ref{Fig4}, which depicts a numerically obtained spectrum with $L_D = 80$ as a function of $\phi_I$. As $\phi_I$ is scanned
from $0$ to $2\pi$, the energy of each subgap state in the deformation region continuously connects to the energy of the next
level. We can also recognize the state at the boundary of QC $I$ with the vacuum [cf.~Fig.~\ref{Fig3}(a)]. Notably, the motion of
the eigenstates with $\phi_{I}$ represents pumping analogues to Laughlin's pumping in the integer quantum Hall
effect~\cite{Laughlin,Zak,ZoharKobi}.

\begin{thebibliography}{32}
\bibitem[{xi()}]{xi}
\bibinfo{note}{Note that the case of $\Delta = 0$ is the return probabilty,
  $\xi_n=|\psi_n|^2$. This generalization is required due to the finite width
  of subgap states.}
\bibitem[{shi()}]{shift}
\bibinfo{note}{To align the energy spectra of the Harper and Fibonacci QC, a
  constant shift is added to the hopping amplitude.}
\end{thebibliography}
\end{document}